\documentclass[twocolumn,showpacs,preprintnumbers,amsmath,amssymb,superscriptaddress,prl]{revtex4}

\usepackage{graphicx}
\usepackage{dcolumn}
\usepackage{bm}
\usepackage{color}
\usepackage{ulem}

\begin{document}

\title{Pressure Tuning of an Ionic Insulator into a Heavy Electron Metal: An Infrared Study of YbS}

\author{M.~Matsunami}
 \altaffiliation[Electronic address: ]{matunami@spring8.or.jp}
 \affiliation{Institute for Solid State Physics, The University of Tokyo, Kashiwa 277-8581, Japan}
\affiliation{Department of Physics, Graduate School of Science, Kobe University, Kobe 657-8501, Japan}

\author{H.~Okamura}
\altaffiliation[Electronic address: ]{okamura@kobe-u.ac.jp}
\affiliation{Department of Physics, Graduate School of Science, Kobe University, Kobe 657-8501, Japan}

\author{A.~Ochiai}
\affiliation{Center for Low Temperature Science, Tohoku University, Sendai 980-8578, Japan}

\author{T.~Nanba}
\affiliation{Department of Physics, Graduate School of Science, Kobe University, Kobe 657-8501, Japan}

\date{\today}

\begin{abstract}
Optical conductivity [$\sigma(\omega)$] of YbS has been measured under pressure up to 20\,GPa. 
Below 8~GPa, $\sigma(\omega)$ is low since YbS is an insulator with an energy gap between fully occupied 4$f$ state and unoccupied conduction ($c$) band. 
Above 8\,GPa, however, $\sigma(\omega)$ increases dramatically, developing a Drude component due to heavy carriers and characteristic infrared peaks. 
It is shown that increasing pressure has caused an energy overlap and hybridization between the $c$ band and 4$f$ state, thus driving the initially ionic and insulating YbS into a correlated metal with heavy carriers. 
\end{abstract}


\maketitle

Physical properties of strongly correlated rare-earth compounds, most typically Ce- or Yb-based intermetallics, have attracted much interest \cite{lawrence}. 
In these compounds, a hybridization between the conduction ($c$) and $f$ electrons leads to a duality between localized and delocalized electronic properties. 
For example, a metallic ground state with large effective mass, often referred to as the heavy electron state, is formed. 
The mass enhancement is due to strong correlation of the $f$ electrons. 
In addition, a non-integer mean valence of rare-earth atom, i.e., an intermediate valence (IV), is observed. 
It results from a mixing of two electronic configurations, and is also a manifestation of the duality. 
Optical conductivity [$\sigma(\omega)$] technique has been a powerful probe for the $c$-$f$ hybridized electronic structures near the Fermi level ($E_{\rm F}$) \cite{degiorgi-review,garner,dordevic,degiorgi,hancock,universal}. 
The dynamics of heavy electrons is observed as a narrow Drude peak in $\sigma(\omega)$ \cite{degiorgi-review}. 
In addition, a characteristic mid-infrared (MIR) peak in $\sigma(\omega)$ has been observed for many Ce and Yb compounds \cite{garner,dordevic,degiorgi,hancock,universal}. 
These features have been analyzed in terms of a $c$-$f$ hybridized band model \cite{degiorgi-review,garner,degiorgi,dordevic,hancock,universal}. 
This model has been shown to explain the basic systematics of the MIR peak energies observed for various compounds with different degrees of hybridization \cite{dordevic,degiorgi,hancock,universal}. 
However, the microscopic nature and formation process of the $c$-$f$ hybridized state are yet to be established.

This Letter addresses the above problem by studying $\sigma(\omega)$ of YbS under external pressure. 
In contrast to the previous $\sigma(\omega)$ studies \cite{dordevic,degiorgi,hancock,universal} where different compounds or chemically alloyed compounds were compared, a pressure study on a given compound has the advantage of tuning the electronic structures without changing other chemical properties, and also without causing an alloy-induced disorder in the crystal lattice. 
This should make it easier to extract essential information on the microscopic electronic structures from the measured $\sigma(\omega)$.

YbS at ambient pressure is an ionic (Yb$^{2+}$S$^{2-}$) insulator with an energy gap of $\sim$1.3\,eV between a fully occupied 4$f$ state and an unoccupied $c$ band \cite{syassen-1,syassen-2,PES}. 
An optical absorption peak due to the energy gap was observed to shift to lower energy with increasing pressure \cite{syassen-1}. 
Volume compression data of YbS indicated that the Yb mean valence deviated from 2 above $\sim$10\,GPa \cite{syassen-1}. 
These results suggested that an IV state was formed above 10\,GPa through a hybridization between the $c$ band and $f$ state, although its microscopic nature was unknown due to a lack of low energy optical data below 0.5\,eV. 
Hence YbS should be a well suited system for studying the $c$-$f$ hybridized electronic structures, where the $c$ band and $f$ state initially separated in energy can be continuously tuned into hybridization by external pressure. 
In addition, the S\,3$p$ valence band is well separated ($\sim$4\,eV below $E_{\rm F}$ \cite{PES}) from the $c$ band and $f$ state. 
Both the chemical composition and the crystal structure (NaCl type) of YbS are quite simple, with the latter stable up to above 20\,GPa \cite{syassen-1}. 
These features should greatly simplify the analysis of $\sigma(\omega)$, making YbS even more attractive.

We have measured the reflectivity [$R(\omega)$] of YbS at photon energies 0.02-1.2\,eV under pressure up to 20\,GPa, and have derived $\sigma(\omega)$. 
This low photon energy range has enabled us to clarify the ground state of YbS under pressure. 
As the energy gap is closed with pressure, both a Drude peak and characteristic IR peaks emerge in $\sigma(\omega)$. 
We show that YbS above 10\,GPa is a heavy electron metal, whose electronic structures can be understood with $c$-$f$ hybridized bands.

The YbS samples used were single crystals grown with the Bridgman method. 
High pressures were generated with a diamond anvil cell (DAC). 
A freshly cleaved surface of a small sample was placed on the culet face of a diamond anvil, together with a gold film used as a reference of reflectivity. 
Either fluorinert or glycerin was used as the pressure transmitting medium, and the pressure was monitored with the ruby fluorescence method. 
Diamond anvils with culet diameters of 0.6 and 0.8\,mm were used in the MIR and far-IR, respectively. 
To accurately perform IR spectroscopy within such a limited sample space, a synchrotron radiation source was used at the beamline BL43IR of SPring-8 \cite{BL43IR}. 
$R(\omega)$ in vacuum was measured from 10\,meV to 30\,eV, then $\sigma(\omega)$ and other optical functions were obtained with the Kramers-Kronig (KK) analysis of $R(\omega)$ \cite{dressel}. 
All the experiments were done at room temperature.

\begin{figure}[t]
\begin{center}
\includegraphics[width=0.43\textwidth,clip]{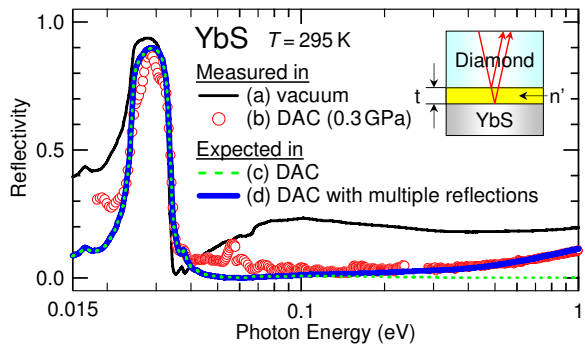}
\caption{
(Color online) (a)-(d) Reflectivity spectra of YbS measured under different conditions. 
The diagram illustrates multiple reflections of the IR ray. 
See the text for details. 
} 
\end{center}
\end{figure}

In Fig.~1, spectrum (a) was measured in vacuum. 
$R(\omega)$ is low except for a high reflectivity band around 30\,meV due to optical phonon. 
Spectrum (b) was measured in a DAC at a low pressure of 0.3\,GPa. 
A portion of the spectrum near 0.3\,eV is not shown due to strong absorption by diamond. 
The difference between (a) and (b) mainly arises from the large refractive index of diamond: the normal-incidence $R(\omega)$ of a material with a real (imaginary) refractive index $n$ ($\kappa$) is expressed as \cite{dressel} 
\begin{equation}
R(\omega)= [(n-n_0)^2+\kappa^2]/[(n+n_0)^2+\kappa^2],
\end{equation}
where $n_0$=1 and 2.4 for vacuum and diamond, respectively. 
Spectrum (c) is the $R(\omega)$ {\it expected} in a DAC, given by Eq.~(1) with $n(\omega)$ and $\kappa(\omega)$ obtained from KK analysis of the spectrum (a). 
Spectrum (b) is seen to deviate from (c) as the energy increases. 
Our analyses have shown that this deviation is due to multiple reflections from a thin layer of pressure medium between the sample and diamond, as illustrated in Fig.~1: 
Spectrum (d) shows a simulation based on such multiple reflections \cite{multiple} with spectrum (a) and $n't$=0.26\,$\mu$m, where $n'$ and $t$ are the refractive index and thickness of the medium, respectively, which can well reproduce spectrum (b) \cite{footnote3}. 
Hereafter $R(\omega)$ relative to diamond will be denoted as $R_d(\omega)$.

\begin{figure}[t]
\begin{center}
\includegraphics[width=0.48\textwidth,clip]{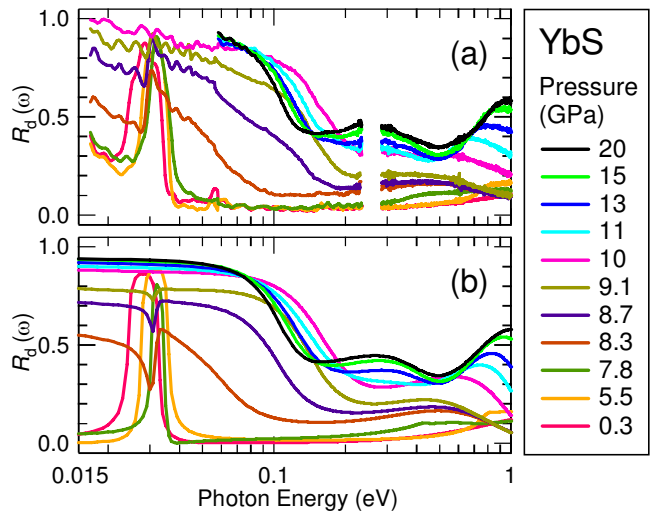}
\caption{
(Color) (a) Reflectivity $R_d(\omega)$ of YbS measured in DAC at pressures up to 20\,GPa. 
(b) Drude-Lorentz fitting to the raw $R_d(\omega)$ data in (a). 
} 
\end{center}
\end{figure}

Figure~2(a) shows the pressure dependence of $R_d(\omega)$. 
(Above 10\,GPa the spectra were measured above 60\,meV only.) 
$R_d(\omega)$ exhibits dramatic changes with pressure above 8\,GPa: 
At 8-10\,GPa, $R_d(\omega)$ below $\sim$0.2\,eV rapidly increases and the phonon peak disappears. 
The high reflectivity below 0.1\,eV is clearly due to plasma reflection, which explicitly demonstrates that {\it the energy gap is closed at $\sim$8\,GPa} and that {\it the ground state above 8\,GPa is metallic}. 
Above 10\,GPa, $R_d(\omega)$ above 0.2\,eV also increases significantly and develops marked structures. 
To obtain $\sigma(\omega)$, Drude-Lorentz spectral fittings \cite{dressel} were performed on the measured $R_d(\omega)$ spectra in Fig.~2(a). 
Effects of the diamond and multiple reflections were taken into account as already described. 
The $R_d(\omega)$ spectra obtained from the fitting are shown in Fig.~2(b). 
$\sigma(\omega)$ calculated from the fitting parameters are shown in Fig.~3(a), together with $\sigma(\omega)$ measured in vacuum. 
An example of the fitting is also shown in Figs.~3(b) and 3(c). 
Below 8\,GPa, $\sigma(\omega)$ shows a peak shifting to lower energy with pressure. 
As previously discussed \cite{syassen-1}, this peak is due to optical absorption across the energy gap and its red shifts show a decrease of gap with pressure. 
Since the $c$ band is mainly derived from the Yb\,5$d$, its bottom is lowered by pressure through the increases in the $c$ bandwidth and 5$d$ crystal field splitting, resulting in the decrease of gap. 
Hereafter this peak will be referred to as the ``gap peak''. 
At 8.3\,GPa the gap peak has almost disappeared, and a rise of $\sigma(\omega)$ toward zero energy emerges, which is a Drude peak due to free carriers. 
At 8-10\,GPa range, the growth of Drude peak is the main spectral development in $\sigma(\omega)$ \cite{footnote}. 
Above 10\,GPa, however, a MIR peak centered at $\sim$0.25\,eV and a near-IR (NIR) peak at 0.7-0.9\,eV grow rapidly with pressure, together with a further growth of the Drude peak.

\begin{figure}[t]
\begin{center}
\includegraphics[width=0.47\textwidth,clip]{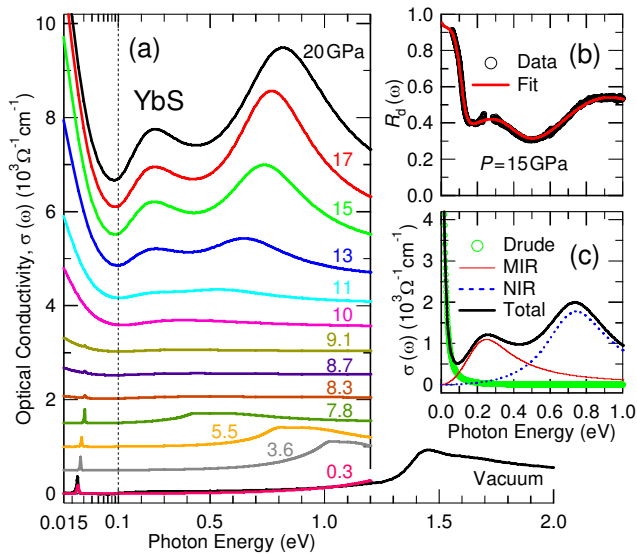}
\caption{
(Color online) (a) $\sigma(\omega)$ of YbS at 0.3-20\,GPa, and that obtained with KK analysis of $R(\omega)$ measured in vacuum. 
The spectra are offset for clarity. 
A logarithmic scale is used below 0.1\,eV. 
(b)(c) Examples of fitting for 15\,GPa data. 
} 
\end{center}
\end{figure}

\begin{figure}[t]
\begin{center}
\includegraphics[width=0.46\textwidth]{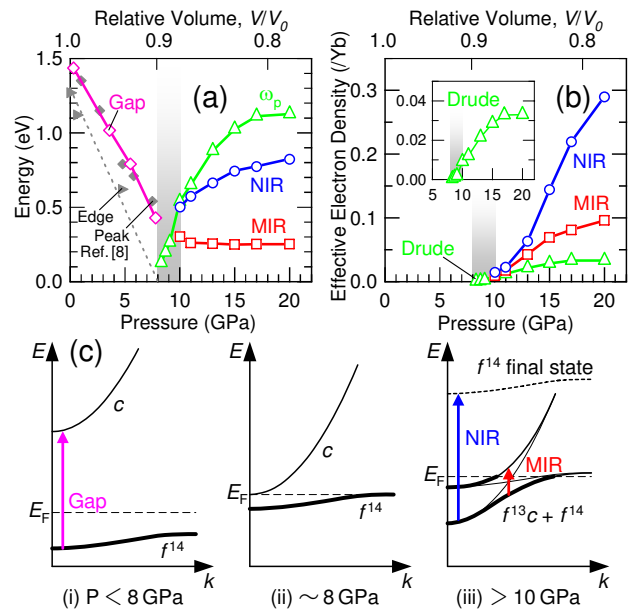}
\caption{
(Color online) 
(a) Pressure dependence of the gap peak, bare plasma frequency ($\omega_p$), MIR and NIR peaks. 
Previous data from Ref.~\cite{syassen-1} are also plotted. 
The dotted lines and the gray area are guide to the eye. 
(b) Pressure dependence of the effective electron density ($N^\ast$) per Yb. 
(c) Schematic illustrations for the electronic structures in YbS. 
} 
\end{center}
\end{figure}

Figure~4(a) shows the observed peak energies and the bare plasma energy ($\omega_p$) versus pressure and the corresponding volume \cite{syassen-1}. 
The absorption edge energies taken from previous work \cite{syassen-1} are also plotted for comparison. 
Both the gap peak and the absorption edge show almost linear shifts with pressure. 
The shift of absorption edge extrapolates to zero at $\sim$8\,GPa, which is consistent with the gap closing at $\sim$8\,GPa indicated by the present data. 
The rapid increase of $\omega_p$ above 8\,GPa corresponds to the emergence of metallic ground state. 
The effective electron density $N^\ast = N/m^\ast$, where $N$ and $m^\ast$ are the density and effective mass of the electrons, can be calculated from the spectral weight in $\sigma(\omega)$ \cite{dressel}. 
In Fig.~4(b), $N^\ast$ is plotted for the Drude, MIR and NIR peaks.

Figure 4(c) shows our model for the electronic structures of YbS under pressure, illustrated in a band picture. 
Well below 8\,GPa, as indicated in (i), Yb is divalent with the fully occupied 4$f$ band. 
The gap peak is due to optical excitations as indicated by the arrow, which can be also regarded as $f^{14} \rightarrow f^{13}c$ excitations in the atomic picture. 
The decrease of energy gap with pressure was already discussed above. 
When the pressure reaches $\sim$8\,GPa, as in (ii), the gap closes as the $c$ band begins to overlap with the $f$ band. 
This leads to a metallic state with free carriers as observed in $R(\omega)$ and $\sigma(\omega)$. 
Then above 10\,GPa, the entire $f$ band has overlapped with the $c$ band, as in (iii). 
In this situation, many electrons have been transferred to the $c$ band, leaving unoccupied $f$ states and hence resulting in an IV. 
A hybridization between the $c$ and $f$ bands should create new channels for optical excitations, as in (iii), resulting in the observed MIR peak above 10\,GPa. 
In fact, such a model of MIR peak based on $c$-$f$ hybridized bands has been successfully used to analyze $\sigma(\omega)$ of many Ce- and Yb-based IV metals at ambient pressure \cite{garner,degiorgi,dordevic,hancock,universal}. 
In the atomic picture, such a hybridized IV state should be expressed as ($f^{13}c + f^{14}$), i.e., a quantum mechanical superposition of the $f^{13}c$ and $f^{14}$ configurations \cite{lawrence}.

In our model, the density of free carriers above 10\,GPa should be given by the density of unoccupied $f$ states, and hence by the increase of Yb mean valence from 2. 
Such a simple analysis should be possible for YbS because other states such as the S\,3$p$ valence band are located away from $E_{\rm F}$. 
In fact, the observed increase of the Drude weight with pressure [inset of Fig.~4(b)] is remarkably similar to that of the Yb mean valence above 10\,GPa \cite{syassen-2}. 
The observed mean valence of 2.4 at 20\,GPa \cite{syassen-2} suggests $N\sim$0.4 per Yb. 
Then the observed value of $N^\ast\sim$0.033/Yb for the Drude peak at 20\,GPa [Fig.~3(b)] should correspond to $m^\ast\sim$12 (in units of the rest electron mass). 
Namely, the narrow Drude response in the IV state of YbS is due to mass-enhanced, heavy carriers. 
This result is consistent with our model (iii) of Fig.~4(c): since $E_{\rm F}$ is located near the $f$-derived, nearly flat portion of the hybridized bands, the resulting carriers should have enhanced $m^\ast$. 
A mass enhancement of 12\,$m_0$ is well comparable to those observed for other Yb-based IV metals \cite{lawrence}, and should result from the strong correlation of $f$ electrons. 
The observation of such a mass-enhanced, narrow Drude response and a MIR peak for pressure-tuned YbS is quite remarkable, since it demonstrates explicitly that they indeed originate from the $c$-$f$ hybridized state. 
In fact, the observed MIR peak energy of $E_{\rm MIR}$=0.25\,eV for YbS is well comparable to those for other Yb-based IV metals, such as YbAl$_3$ (0.25\,eV \cite{YbAl3}), YbInCu$_4$ (0.25\,eV \cite{hancock,YbInCu4}) and YbAl$_2$ (0.3\,eV \cite{universal}).

The NIR peak, which appears simultaneously with MIR peak in $\sigma(\omega)$, is likely to result from optical excitation from the $f^{13}c$ component of the IV state to the (pure) $f^{14}$ state. 
Note that such a process becomes possible only after the formation of IV state. 
In the band picture, it may be expressed as shown in (iii) of Fig.~4(c). 
The growth of NIR peak with pressure, as observed in Fig.~4(b), should correspond to the increase of $f^{13}c$ component in the IV state, which is consistent with the observed increase of mean valence \cite{syassen-1,syassen-2}. 
The previous optical work on YbS also proposed a similar mechanism for the NIR peak \cite{syassen-1} based on data above 0.5\,eV. 
However, the present work with lower energy data provides stronger evidence, as both MIR and NIR peaks can be well explained on the basis of $c$-$f$ hybridized state. 
A similar NIR peak was also observed for Yb(Ag,In)Cu$_4$, and was also attributed to $f^{14}$-related excitations \cite{hancock}.

Now we shall consider more microscopically the nature of $c$-$f$ hybridized state in YbS with respect to pressure tuning, which is the novel feature of this work. 
On the basis of a periodic Anderson Hamiltonian, the $c$-$f$ hybridization energy $V^\ast$, renormalized from its bare value $V$ by $f$ electron correlation, can be expressed as \cite{cox} 
\begin{equation}
V^\ast \propto \sqrt{T_{\rm K} W}, 
\end{equation}
where $T_{\rm K}$ is the characteristic Kondo temperature and $W$ is the $c$ bandwidth in the absence of hybridization. 
The $E_{\rm MIR}$ of IV metals has been interpreted in terms of $V^\ast$ \cite{garner}, and in fact a universal relation has been found between the measured $E_{\rm MIR}$ and estimated $\sqrt{T_{\rm K} W}$ for many IV compounds \cite{dordevic,hancock,universal}. 
$E_{\rm MIR}$ of YbS plotted in Fig.~4(a), in contrast, shows only weak pressure dependence, suggesting that $V^\ast$ is nearly unchanged by pressure. 
This may appear puzzling since it is usually assumed that an external pressure causes a stronger hybridization. 
However, while the bare $V$ should be certainly increased by pressure, the renormalized $V^\ast$ may not be necessarily so, since it results from complicated correlation. 
For YbS, $W$ in Eq.~(2) is the width of (5$d$-6$s$)-derived $c$ band, and should be increased with pressure, since the interatomic distance decreases. 
Therefore, the observed pressure-independent $V^\ast$ and Eq.~(2) suggest that $T_{\rm K}$ decreases with pressure in YbS, and hence, that the $f$ electrons (or equivalently $f$ holes) in the metallic phase attain more localized characters with increasing pressure. 
These are consistent with the increase of mean valence toward 3. 
A decrease of $T_{\rm K}$ with pressure, contrary to an increase widely observed for Ce compounds, has also been reported for other Yb compounds \cite{theory}. 
Theoretical analysis on Yb compounds has suggested that competing effects related with the hybridization and ionic radius may cause a complicated pressure dependence of $T_{\rm K}$ \cite{theory}. 
Further study including a measurement of $T_{\rm K}$ is needed to further clarify the origin of the pressure-independent $V^\ast$ in the context of Kondo physics for YbS.

In conclusion, $\sigma(\omega)$ of YbS has been measured as it is tuned by pressure from an ionic insulator into a metal. 
A narrow Drude peak due to heavy carriers and characteristic IR peaks have been found in $\sigma(\omega)$, strongly suggesting that YbS above 10\,GPa is a correlated heavy electron metal with an IV. 
A model based on $c$-$f$ hybridized bands has been used to analyze the pressure evolution of the spectra. 
A decrease of $T_{\rm K}$ with pressure has been suggested by the analysis of $\sigma(\omega)$, which indicates that the $f$ electrons (holes) in the metallic phase attain more localized character with increasing pressure.

We thank M.~Taguchi for fruitful discussion, and Y.~Ikemoto and T.~Moriwaki for technical assistance. 
M.~M. would like to thank S.~Shin and A.~Chainani for their encouragement, and H.~O. acknowledges financial support from MEXT (Innovative Area ``Heavy Fermion'' 21102512-A01) and JSPS (Scientific Research B 17340096). 
Work at SPring-8 was approved by JASRI (2005B0621-NSa-np, 2006A1186-NSa-np, 2007B1314), and that at UVSOR by the Joint Studies Program of IMS.



\begin{thebibliography}{99}
\bibitem{footnote3}
The higher baseline of (b) than (c) below 0.1\,eV is due to reflection from the rear surface of the sample.
\bibitem{footnote} 
To well reproduce $R(\omega)$, a Lorentz oscillator centered at $\sim$0.2\,eV was used in the fitting, which was much weaker than the MIR and NIR peaks.
\end{thebibliography}
\end{document}